%% file: pks2155ms.tex
\documentclass[12pt,preprint]{aastex}

\begin{document}

\title{New Multiwavelength Observations of PKS 2155--304 and Implications for the Coordinated Variability Patterns of Blazars}

\shorttitle{PKS 2155--304; Analysis of Multiwavelength Variations}
\shortauthors{Osterman et al.}

\author{M. Angela Osterman\altaffilmark{a,b}, H. Richard Miller\altaffilmark{b}, Kevin Marshall\altaffilmark{b}, Wesley T. Ryle\altaffilmark{b}, Hugh Aller\altaffilmark{c}, Margo Aller\altaffilmark{c}, John P. McFarland\altaffilmark{d} }

\altaffiltext{a}{Egan Observatory, Florida Gulf Coast University, 10501 FGCU Blvd., Fort Myers, FL, 33965-6565; mosterma@fgcu.edu}

\altaffiltext{b}{Dept. of Physics and Astronomy, Georgia State University, 1 Park Place Suite 730, Atlanta, GA, 30302; osterman@chara.gsu.edu}

\altaffiltext{c}{University of Michigan, Dept. of Astronomy, 500 Church St. 830 Dennison, Ann Arbor, MI, 48109-1042}

\altaffiltext{d}{Kapteyn Instituut, Rijksuniversiteit Groningen, 9747 AD Groningen, The Netherlands}

\begin{abstract}
The TeV blazar PKS 2155--304 was the subject of an intensive 2 week optical and near-infrared observing campaign in 2004 August with the CTIO 0.9m telescope.
During this time, simultaneous X-ray data from \emph{RXTE} were also obtained.
We compare the results of our observations to the results from two previous simultaneous multiwavelength campaigns on PKS 2155--304. 
We conclude that the correlation between the X-ray and UV/optical variability is strongest and the time lag is shortest (only a few hours) when the object is brightest.
As the object becomes fainter, the correlations are weaker and the lags longer, increasing to a few days.
Based on the results of four campaigns, we find evidence for a linear relationship between the mean optical brightness and lag time of X-ray and UV/optical events.
Furthermore, we assert that this behavior, along with the different multiwavelength flare lag times across different flux states is consistent with a highly relativistic shock propagating down the jet producing the flares observed during a high state.
In a quiescent state, the variability is likely to be due to a number of factors including both the jet and contributions outside of the jet, such as the accretion disk.
\end{abstract}


\section{Introduction}

In the broad class of extragalactic objects known as active galactic nuclei (AGNs), blazars distinguish themselves in many ways. 
Blazars are a radio-loud subclass of AGNs, typically featuring very core-dominated radio morphologies. 
This class of objects is characterized by (1) a strong, featureless continuum, (2) strong (up to $\sim$20\%) and variable polarization, and (3) the presence of large-amplitude, rapid flux variability observed on timescales of less than an hour to several years at all wavelengths.
The most extreme and unique property of blazars is their highly beamed continuum, most likely produced by a jet of relativistic material aimed close to the observer's line of sight. 
BL Lac objects have particularly weak spectral lines, so their redshifts, and hence luminosities, are difficult to determine.
Blazars exhibit the most extreme variability observed for any class of AGN.
See \citet{urr95} and the references therein for further details.

The spectral energy distribution (SED) of blazars is distinguished by two peaks: one in the radio/UV range and the other in the X-ray/$\gamma$-ray range. 
The spectrum in the radio to UV range is generally agreed to arise from synchrotron emission from relativistic electrons spiralling around the jet's magnetic field lines \citep{ulr97}. 
The X-ray/$\gamma$-ray spectrum is most likely due to inverse Compton (IC) emission in low-frequency peaked BL Lac objects (LBLs) and flat-spectrum radio quasars (FSRQs), while most of the X-rays are the high-energy tail of the synchrotron emission in high-frequency peaked BL Lac objects (HBLs) \citep{ulr97}. 
Any $\gamma$-ray emission is most likely explained as inverse Compton scattering of photons by relativistic electrons in the jet (e.g. \citet{mar85}).

Multiwavelength campaigns are a very useful tool for investigating the underlying structure and physics of blazar jets.
Assuming different areas of the jet emit different wavelengths of light, simultaneous multiwavelength campaigns can examine the physical processes operating in the jet and indicate spatial relationships between different emission regions. 
Correlations in multiwavelength flares can be used to constrain models predicting the source of photons which are upscattered to produce the IC emission observed in blazars.
In many blazars, the X-ray variability is observed to be more rapid and of larger amplitude than the variability at longer wavelengths, implying that this emission originates from a smaller region than the optical and radio emitting regions.
We must understand the physics of the region near the central engine in order to fully understand blazars, since this is where jet particles are collimated and accelerated to relativistic speeds. 

PKS 2155--304 is one of the best-studied blazars in the southern hemisphere.
It was first observed in the radio as part of the Parkes survey \citep{shi74}.
\citet{gri78} noted a blue stellar object with a featureless continuum at the same location.
\citet{sch79} and \citet{hew80} first identified PKS 2155--304 as a BL Lac object.
\citet{sch79} report the first X-ray observations of this object, using \emph{HEAO 1}.
Significant optical variability ($\sim$1.5 mag) and polarization ($\sim$5\%) was observed by \citet{gri79}.
A redshift of 0.117 $\pm$ 0.002 was determined by \citet{bow84}; the current accepted value is 0.116 \citep[e.g., ][]{fal93}.
The long-term R-band light-curve is displayed in Figure 1.
Recently, it has been detected at TeV energies \citep{rob99,dja03,emm06}, thus opening up new possibilities for multiwavelength campaigns.
During the past 15 years, PKS 2155--304 has been the subject of numerous campaigns examining possible correlations between X-ray, UV, optical, and radio variability.

The two most prominent and intensive of these earlier campaigns occurred in 1991 November \citep{ede95, urr96} and 1994 May \citep{urr96, urr97}.
In both campaigns, at least one set of temporally correlated flares was observed in X-ray and ultraviolet wave bands.
The 1991 campaign showed correlated optical flares as well.
However, the time lags between flares in different wave bands differed in each campaign.

Since PKS 2155--304 is a HBL, the radio through X-ray emission are all thought to be due to the synchrotron process in the jet, with higher energy emission occuring closer to the central engine than lower energy emission \citep{ulr97}.
Hence, flares at lower energies should lag flares at higher energies.
The flaring activity in both the 1994 and 1991 campaigns follow this trend.
The results from these campaigns will be discussed further in conjunction with the results from the new 2004 campaign on PKS 2155--304.

\section{Observations Summary and Data Reduction}

PKS 2155--304 was the subject of an intense 10 night optical campaign performed at the Cerro Tololo Inter-American Observatory (CTIO) 0.9 m telescope in 2004 August.
During this time, X-ray observations were also obtained with the Rossi X-ray Timing Explorer (\emph{RXTE}).
Integrations of a few hundred to a few thousand seconds were performed several times a day for 21 days with \emph{RXTE}'s Proportional Counter Array (PCA).
These integrations were followed by a second set of longer (a few to several ks) \emph{RXTE} integrations.
These observations spanned 6 days in 2004 September.
In mid-August, high time resolution optical observations were obtained in the R-band in order to search for the presence of microvariability.
In August, September, and October, additional B-, V-, and R-band observations were obtained through the Small and Moderate Aperture Research Telescope System (SMARTS) consortium 1.3 m telescope and with the 1.8 m Perkins Telescope at Lowell Observatory.

Radio observations at 14.5 and 8.0 GHz were performed at the University of Michigan's Radio Astronomy Observatory (UMRAO).
Most of the 14.5 GHz observations coincide with the \emph{RXTE} observations.
However the 14.5 GHz coverage is much sparser than the \emph{RXTE} observations.
These radio observations are distributed over the months of August, September, and October.
The 8.0 GHz observations, which were also much sparser in coverage than those of \emph{RXTE}, all took place in September.
A summary of all observations taken for this campaign is included in Table 1.

The \emph{RXTE} observations show three clearly defined flaring events, two during August and one during September.
The optical observations show two clearly defined flares in August and elevated brightness states in September and October.
The radio observations do not include any well-defined flaring events, but the 14.5 GHz data show a general increase in flux from August to mid-October and the 8.0 GHz data indicate an increase in flux during September.
The multiwavelength lightcurves from this campaign are displayed in Figure 2.

\subsection{X-ray Data Reduction}

The \emph{RXTE} X-ray light curve was extracted using the FTOOLS version 5.2 software package.
During nearly all of our observations, PCUs 1, 3, and 4 were turned off.
Therefore, data were only extracted from PCUs 0 and 2.
Despite the loss of the propane layer onboard PCU 0 during May of 2000, the signal-to-noise ratio was much greater when using data from both PCUs 0 and 2.
To further enhance the signal-to-noise ratio, only data from layer 1 of the PCA were analyzed.
No data from the HEXTE cluster or the other PCUs were used.
All of the data analyzed here were taken while the spacecraft was in STANDARD-2 data mode.

Data were extracted only when the target's Earth elevation angle was $>10\deg$, pointing offset $<0.02\deg$, PCUs 0 and 2 both on, the spacecraft more than 30 minutes after SAA passage, and electron noise less than 0.1 units.
Since the background response of the PCU is not well defined above 20~keV, only channels $0-44$ ($2-20$~keV) are included in this analysis.

Because the PCA is a non-imaging detector, background issues can be critically important during analysis.
The faint-mode ``L7'' model, developed by the PCA team, was used here.
This model provides adequate background estimation for objects with less than 40 counts $s^{-1}$.
Background files were extracted using {\tt pcabackest}~version 3.0.

\subsection{Optical Data Reduction}

The Lowell Observatory and CTIO (excluding SMARTS) observations were taken, reduced, and processed by the Program for Extragalactic Astronomy (PEGA) group at Georgia State University. 
Observations from Lowell Observatory utilized the PRISM camera on the 1.8 m Perkins Telescope and Johnson color \emph{BVRI} filters.
Observations from the CTIO 0.9 m utilized a SITe CCD camera equipped with Johnson \emph{BV} and Cousins $R_{c}I_{c}$ filters.
Bias/zero and flat-calibration frames were taken along with the PKS 2155--304 object frames.
Dark calibration frames were not required for the CCDs on these telescopes since each chip contains a dark pixel strip. 
All data reduction utilized standard NOAO IRAF\footnote{IRAF is distributed by the National Optical Astronomy Observatories, which are operated by the Association of the Universities for Research in Astronomy, Inc., under cooperative agreement with the National Science Foundation.}  routines including {\tt ccdproc}, {\tt flatcombine}, and {\tt zerocombine}. 
All data processing and 7" aperture photometry were done using the {\tt ccdphot} routine.
The frames provided by the SMARTS program were observed using the CTIO 1.3 m telescope equipped with a Fairchild 447 camera and Johnson \emph{BVRI} filters.
The SMARTS frames were reduced and photometered using methods similar to those used for PEGA data reduction.

\subsection{Radio Data Reduction}

The University of Michigan radio data were obtained using a 26 m prime focus paraboloid equipped with transistor-based radiometers operating at central frequencies of 4.8, 8.0, and 14.5 GHz and room-temperature wide-band High Electron Mobility Pseudomorphic Transistor (HEMPT) amplifiers (with a width of $\sim$10\% of the observing frequency).
Measurements at all three frequencies utilized rotating, dual-horn polarimeter feed systems, which permitted both total flux density and linear polarization to be measured. 
An on-off observing technique was used at 4.8 GHz, and an on-on technique at the other two frequencies.
A typical observation consisted of 8--16 individual measurements over a 25--45 minute period (depending on frequency).
A source selected from a grid of calibrators was observed every 1--2 hr. 
The flux scale was set by observations of Cassiopeia A. 
Details of the calibration and analysis techniques are given in \citet{all85}.

\section{Discussion}

Figure 2 displays the multiwavelength light-curves from this campaign.
The densely sampled X-ray data display two substantial, well-defined flares in August and another flare in September.
The two August flares appear to each have the same duration, about 4 days.
The first flare represents almost a 150\% increase in flux from the level preceding the flares.
The second flare has a somewhat higher peak than the first.
The first flare's peak is narrower than the second and occurs on August 12, about 4 days before the second flare's peak.
The low flux following the second flare shows a steady decline over the month of August.
There is evidence of an elevated flux state, possibly suggesting another smaller flare immediately preceding the large August flare.
The September X-ray data, although not as dense as the August data, suggest that a flare of approximately 6 days duration occurred during that time.
This flare's peak is poorly-defined, but the observations indicate that the September flare's peak flux is at least as high as that observed in the August flares.
The September data also indicate that a smaller flaring event may have occurred immediately preceding the larger flare.

The densely-sampled optical light-curve from CTIO, shown in Figure 3, exhibit two distinct flares in August.
The flares are clearly present in all four optical wave bands, and appear to occur simultaneously.
As is expected, the flux increases more steeply in \emph{B} than in \emph{V}, and more steeply in \emph{V} than in \emph{R}.
It is not entirely clear when these flares begin and end due to the absence of data extending prior to and following this campaign, but the first flare's peak is clearly observed on August 14 and the second flare is clearly flattening by the time our observations end on August 23.
Based on our observations, the first flare has a duration of at least 3 days, the second of at least 8 days.

The observed X-ray flares do not appear to be related to any of the observed optical flares.
There are no clear similarities in the patterns, structure, or timescales of the X-ray flares in relation to the optical flares, as illustrated in Figure 3.
It should be noted, however, that there are many extended periods during which there are either no X-ray observations to compare with optical observations, or vice versa.
It is possible that more flares might be present, but undetected.
A discrete correlation function (DCF) was performed on the simultaneously obtained R-band and X-ray data \citep{ede88}.
No significant correlations between these two wave bands were apparent in the results of this analysis.
The flux observed at 8.0 and 14.5 GHz appears to steadily increase over the course of the UMRAO observations.
This observed increase in radio flux may be related to the smooth increase in optical brightness observed in late August and early September.

\section{Multiwavelength Synchrotron Behavior}

In the campaigns on PKS 2155--304 included in the present study, distinct flares were observed at optical/UV and X-ray wavelengths.
In the 2004 campaign, these flares occurred within several days of each other, but do not appear to be correlated.
However, correlated multiwavelength flares were observed in this object in two previous campaigns when the object was observed in higher flux states (see Table 2).
The intermediate-flux state exhibited flares lagging each other by a few days from X-ray to UV wavelengths \citep{urr97}, while the highest observed state featured flares lagging each other only by $2-3$ hours \citep{ede95}.
In all cases, the X-ray flares lead the lower energy flares.
See Figures 4 and 5 for multiwavelength observations from the 1991 and 1994 campaigns.

In every campaign, the flares observed at higher energies have larger amplitudes than those at lower energies.
In 1991, with the object in a high-flux state, flaring activity between the two \emph{International Ultraviolet Explorer (IUE)} wave bands was observed to be simultaneous down to the limiting observational timescale of $<1-2$ hours \citep{ede95}.
As illustrated in Figure 4, the X-ray, UV, and optical flares all had very similar structure and amplitudes.
In 1994, the object was in a lower flux state.
Flaring activity from X-ray through UV regimes was still temporally correlated, but on significantly longer timescales.
The time lag from the X-ray to UV flares increased to about 2 days \citep{urr97}.
Also, as illustrated in Figure 5, significant differences appeared in the light-curves from each regime.
The longer wavelength flares are significantly broader than the X-ray flare, and exhibit smaller amplitudes.
Small differences in flare structure between the two \emph{IUE} wave bands appeared, as shown in Figure 5, although the flare structures remained quite similar \citep{urr97}.
In 2004, the object was in a still lower flux state, and no UV data were available.
However, the intense simultaneous optical and UV data obtained in 1991 (see Fig. 4) demonstrate that the optical behavior is probably a good indicator of the UV behavior.
The intensive 2004 data show highly correlated optical multiwavelength behavior, with the object becoming somewhat bluer when brighter.
The X-ray flaring, compared to the optical flaring in Figure 3, is of markedly different structure, amplitude, and timescale.
The high-energy flux more than triples, while the optical flux changes by less than a factor of 2.
Both X-ray flares are of similar amplitude, while the second optical flare is significantly stronger than the first.
The X-ray flare peaks occur within 4 days of each other, while the two optical flare peaks are separated by at least 9 days.
No multiday or intraday similarities in X-ray/optical flare structures are present.
It is possible that optical flares resembling the X-ray flares did occur, but no clear correlations are present in the observed optical and X-ray activity.

\section{Suggested Causes of Multiwavelength Flares}

These comparisons of PKS 2155--304 multiwavelength campaigns suggest that the correlation between the X-ray and UV/optical continuum is strongest and the time lags shortest when the object is brightest.
The correlations weaken and the time lags increase when the object is fainter, and in the faintest states the correlations appear to vanish entirely.
The decrease in time lags suggests that the X-ray and optical emission regions are becoming closer spatially, but this does not fully explain the observed behavior.
The similarity in multiwavelength flare structure observed in the higher flux states supports the hypothesis that the X-ray and optical flares are caused by a shock propagating down the jet, stimulating longer wavelengths of synchrotron radiation as the shock moves from the central engine outward.

It can thus be argued that during a high flux state, the shocks are highly relativistic and quickly propagate between X-ray and optical emission regions in the jet.
In an intermediate state, the X-ray and optical emitting regions are further apart, and/or the shock moves more slowly, and the multiwavelength flares show time lags and differences in structure.
In a low state, the X-ray and optical emitting regions are still further apart, and it is possible that some of the flaring activity is not due to jet emission.
It may be that a strong flaring component is present in the jet during a high state, while in a low state a quiescent component independent of the jet is responsible for observed variations.

The behavior observed during a 2003 campaign agrees with our predictions.
\citet{aha05} performed a simultaneous $\gamma$-ray, X-ray, and optical campaign on this object with the High Energy Stereoscopic System (HESS), \emph{RXTE}, and the Robotic Optical Transient Source Experiment (ROTSE), respectively.
Long-term optical observations of PKS 2155--304 indicate that it was in a particularly low flux state during this time (Fig. 1 of this work; Aharonian et al. 2004).
Following the predictions of this work, X-ray and optical observations of this object would not be expected to display significant correlations during this low-flux state.
The simultaneous \emph{RXTE} and ROTSE observations covered 5 days in 2003 October and 6 days in November \citep{aha05}.
During these time periods, \citet{aha05} did not observe any clear correlation between the X-ray and optical observations, suggesting that any correlated activity in these two wave bands would be observed with a lag of $>$6 days.
It should be noted that the sampling rates of X-ray and optical data differ, particularly in the 2003 November observations, making correlation analysis difficult.

In 2006 July, \citet{fos07} performed simultaneous X-ray, UV, and optical observations on this object.
According to the long-term optical data displayed in Figure 1, PKS 2155--304 was at a flux state similar to that observed in 1994 May, i.e. an intermediate state. 
In this state, we predict that somewhat correlated X-ray and UV flares will be observed one to a few days apart, the X-ray flare will be of larger amplitude than the UV flare, the UV and optical activity will be well correlated, and the high-frequency activity will always lead that at lower frequencies.
The observations and results published by \citet{fos07} do indeed show correlated optical and UV activity, as displayed in their Figure 1.
These results also include an X-ray flare reflecting a change in flux by a factor of 5.
This X-ray flare precedes a smaller UV flare by about 1 day, with the UV flux only increasing by about a factor of 1.5, as displayed in \citet{fos07} Fig. 2.

Figure 6 displays the relationship between optical brightness and lag time between correlated X-ray and UV/optical events observed in the 1991, 1994, and 2006 campaigns. 
The campaigns for which correlated events were observed suggest there is a linear relationship between brightness and lag time during intermediate and high states of PKS 2155--304. 
Future campaigns with longer temporal coverage could make it possible to observe correlated X-ray and UV/optical events with lag times of several days to a few weeks, and to better examine the relationship between brightness and lag time.

Figure 6 also displays the observed relationship between the average X-ray flux and lag time. 
Unlike the optical brightness, the X-ray flux does not appear to be well-correlated with the X-ray to UV/optical lag time.
The X-ray and optical brightnesses appear to be loosely correlated over the five campaigns studied.
As displayed in Figure 7, a brighter X-ray state generally corresponds to a brighter optical state, but a given X-ray brightness can have a range of simultaneously observed optical brightnesses.
It is possible that the UV flux behavior correlates with lag time. 
For the 1991 and 1994 campaigns, the only campaigns for which \emph{IUE} data are available, the lower average observed UV flux corresponds to the longer X-ray to UV/optical lag time.
Investigating correlations between UV and other flux regimes and lag time would be a good subject for future campaigns

We also investigated the optical to X-ray spectral shape during each campaign.
To do this, we calculated $\alpha_{OX}$, defined as the ratio of the average observed X-ray flux to the average observed optical flux (see Table 2). 
Even though the X-ray and optical frequencies observed during each campaign differ, comparisons can still be made between significant increases and decreases in $\alpha_{OX}$.
Comparisons between $\alpha_{OX}$ and X-ray brightness, \emph{R} magnitude, and X-ray to UV/optical lag times are plotted in Figure 8.
The changes in spectral shape appear to correlate well with X-ray brightness for the brighter observed X-ray states. 
The spectral shape does not appear to correlate well with optical brightness or lag time. 
\citet{dol07} investigate changes in IR regime spectral shape versus optical brightness in PKS 2155--304.
They also find no clear correlation between spectral shape and optical brightness.

\section{Conclusions}

We present the comparison of our observations to the results from two previous simultaneous multiwavelength campaigns on PKS 2155--304. 
We conclude that the correlation between the X-ray and UV/optical variability is strongest and the lag is shortest when the object is brightest.
As the object becomes fainter, the correlations are weaker and the are lags longer.
Furthermore, we assert that the different lag times across different flux states indicate that a relativistic shock propagating down the jet causes the flares during a high state.
During a low state, the flaring could be due to something other than the jet, for example the accretion disk (e.g. Mangalam \& Wiita 1993), explaining the observed lack of clearly correlated multiwavelength activity in 2004.
These differences in correlated flaring activity could also come from a ``swinging'' jet which has a constant Lorentz factor, but small changes over time in the angle between the jet and the observer \citep{gop92}.
All of the observed flaring activity was consistent with the synchrotron process.
An interesting future prospect would be to conduct more simultaneous multiwavelength campaigns on other blazars to see if similar patterns emerge.

During a recent campaign on fellow XBL PG 1553+11, clear flares were observed at X-ray and optical wavelengths \citep{ost06}.
The peaks of these flares appeared to be separated by about 10 days.
These flares may be correlated, with the X-ray event leading the optical.
If, as has been argued, PKS 2155--304 and PG 1553+11 are inherently similar objects, then the flux state of PG 1553+11 may also be related to the correlation of multiwavelength flares.
Following the patterns observed in PKS 2155--304, a lag between X-ray and optical flares of as long as 10 days would indicate a relatively low flux state in PG 1553+11 during the 2003 campaign.
The long-term R-band data on PKS 2155--304 and PG 1553+11 agree with these assessments of the relative flux states of each object at the observed epochs.
PG 1553+11 was observed around R$=$13.5 to 13.7 during the 2003 campaign, a low state compared to almost all other observations from 2000--2006 \citep{awc04}.
The longterm PKS 2155--304 light-curve also indicates that this object was brighter during the 1991 campaign than in the 1994 campaign, and the 1994 state was brighter than the state during the 2004 campaign.

Additional simultaneous campaigns are needed on other blazars to see if similar patterns emerge in the multiwavelength flaring activity.
Also, future campaigns should be performed on PKS 2155--304 and PG 1553+11 during different flux states to see if the pattern found for PKS 2155--304 continues to be observed in these objects. 
These campaigns would benefit from broader wavelength coverage, particularly in the TeV regime, which would include IC emission from PKS 2155--304, PG 1553+11, and similar objects.
The \emph{GLAST} observatory, scheduled to launch at the end of 2007, would provide valuable $\gamma$-ray regime coverage to similar campaigns on LBLs and FSRQs.
Subsequent campaigns would also benefit from more extensive temporal coverage.

\acknowledgements

M.A.O., H.R.M., K.M., and W.T.R. are supported in part by the Program for Extragalactic Astronomy's (PEGA) Research Program Enhancement funds from Georgia State Universiy (GSU), and by a grant from NASA (NAG5-13733). 
The CTIO observations were obtained through NOAO proposal 04B-0239 submitted by M.A.O. and H.R.M.
The \emph{RXTE} observations were obtained from the HEASARC online public archives.
M.A.O., H.R.M., K.M., J.P.M., and W.T.R. thank Lowell Observatory for generous allocations of observing time on the Perkins telescope with the PRISM camera.
The authors thank Amy Williams-Campbell and PEGA at GSU for the long-term PKS 2155--304 and PG 1553+11 data.
The authors thank the SMARTS consortium for contributing observations to this work.
The UMRAO facility is partially supported by a series of grants from the NSF, most recently NSF-0607523, and by the University of Michigan.

\clearpage
\input{tab1.tex}

\input{tab2.tex}

\clearpage

\begin{figure}
\includegraphics[angle=270, scale=0.65]{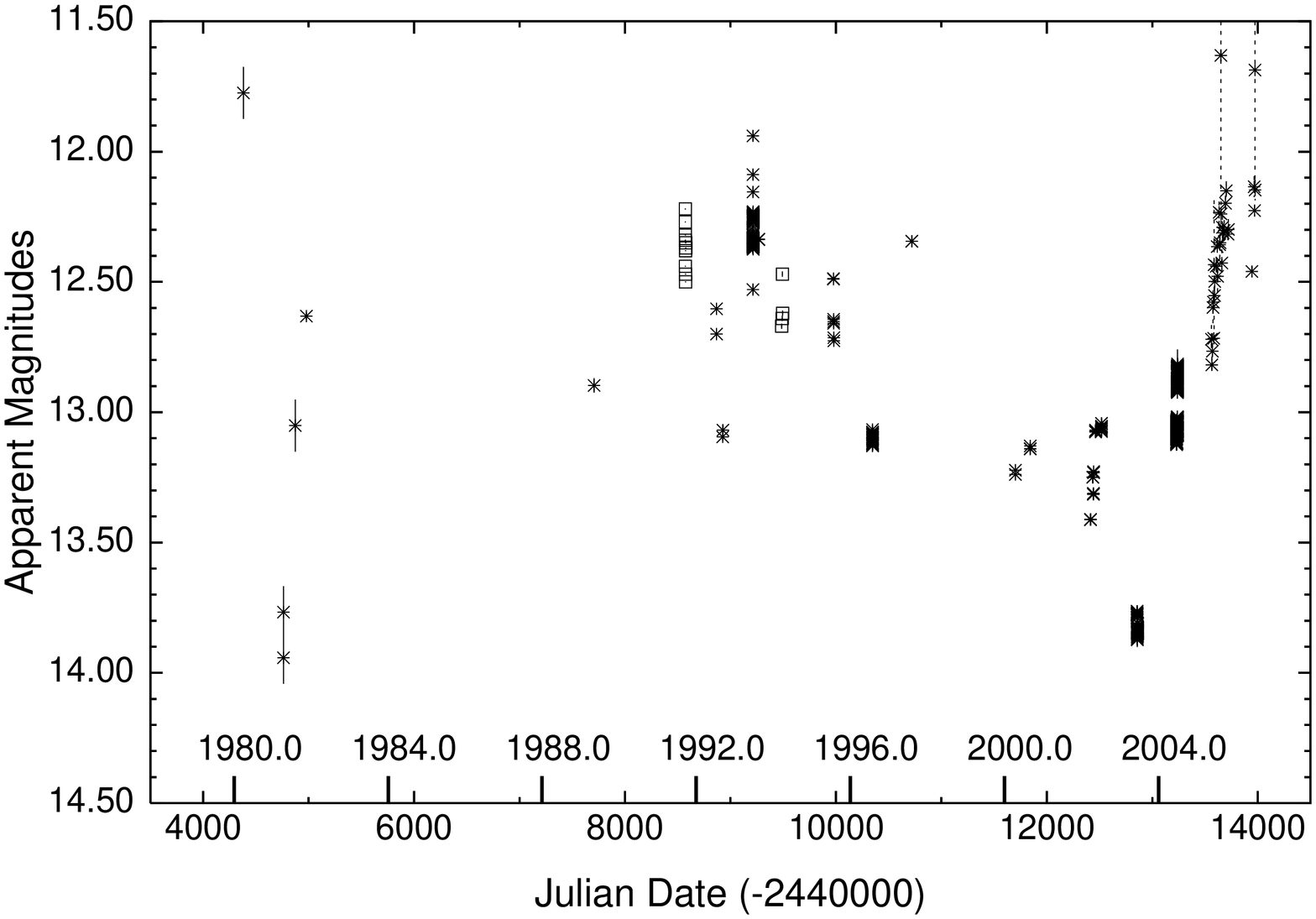}
\figcaption{The long-term R-band light-curve of PKS 2155--304. Data from 1980 to 2003 are published in \citet{awc04}. Open squares are R$_{C}$-band observations from 1991 \citep{cou95} and 1994 \citep{pes97}.}
\end{figure}

\begin{figure}
\includegraphics[angle=270, scale=0.65]{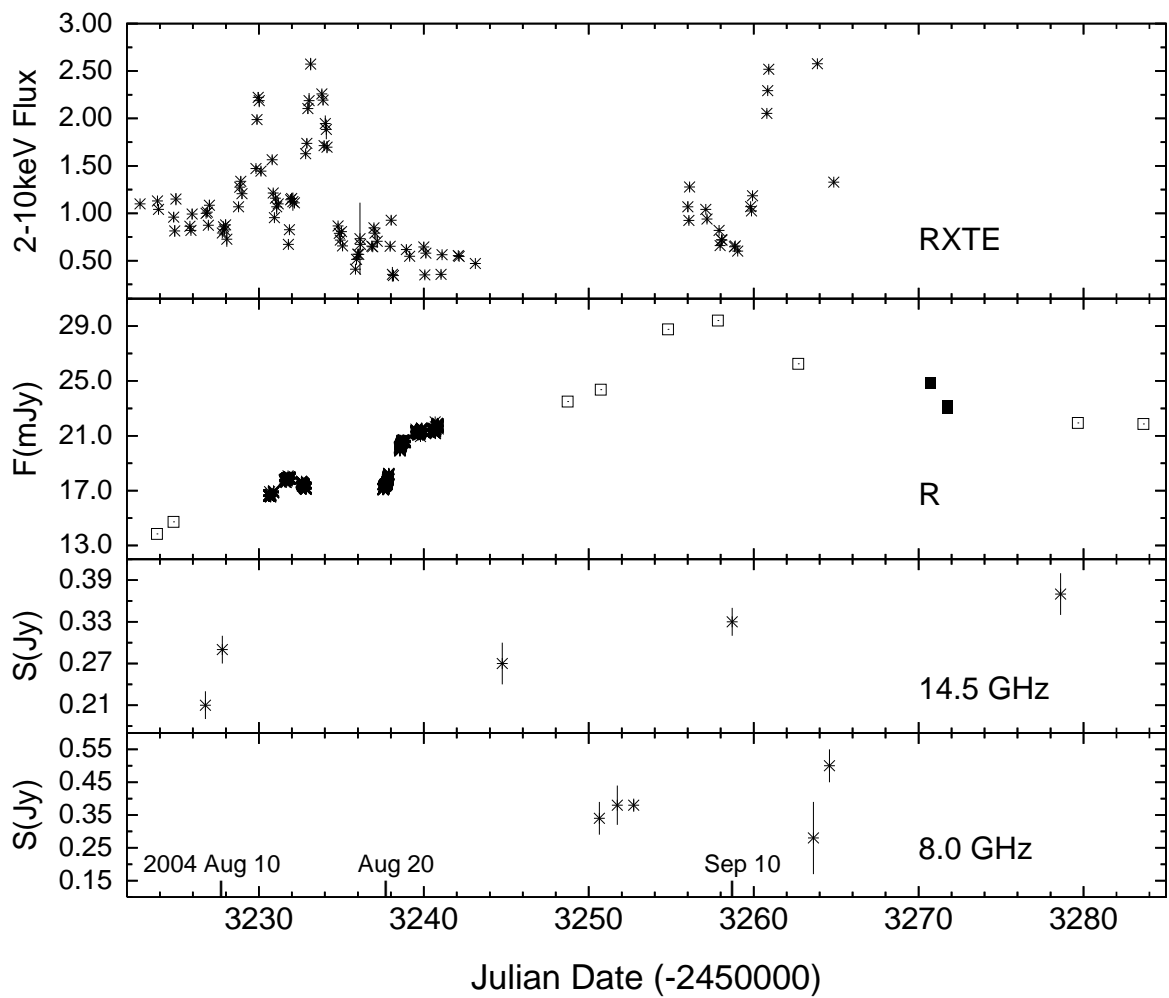}
\figcaption{Simultaneous multiwavelength data taken during the 2004 PKS 2155--304 campaign. 
For the X-ray and optical observations, the error bars are smaller than the plotted points. 
The \emph{RXTE} fluxes are given in units of 10$^{-11}$ erg cm$^{-2}$ sec$^{-1}$. 
For the optical R-band data points, asterisks indicate observations from CTIO, open squares indicate observations from SMARTS, and filled squares indicate observations from Lowell Observatory. 
The radio observations were all obtained from UMRAO.}
\end{figure}

\begin{figure}
\includegraphics[angle=270, scale=0.65]{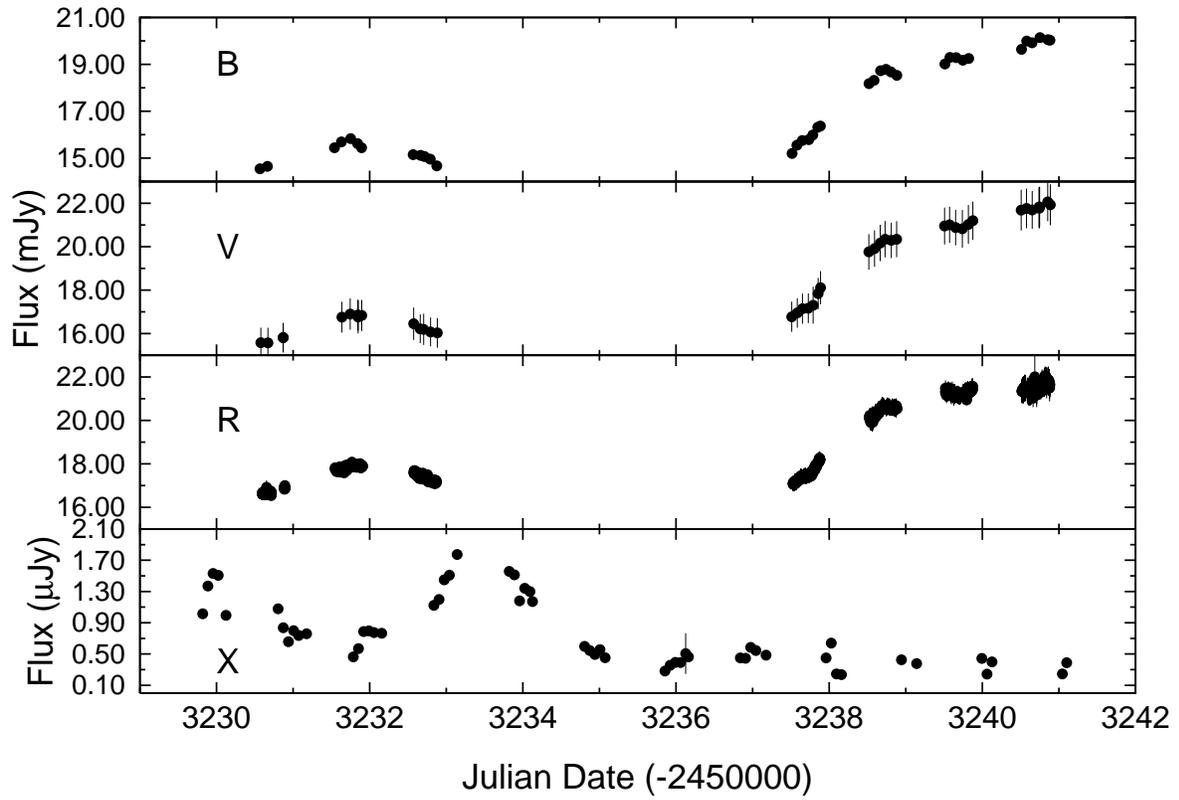}
\figcaption{PKS 2155--304 BVR and X-ray flux data from 2004 August. Error bars may be smaller than the plotted points.}
\end{figure}

\begin{figure}
\includegraphics[angle=270, scale=0.60]{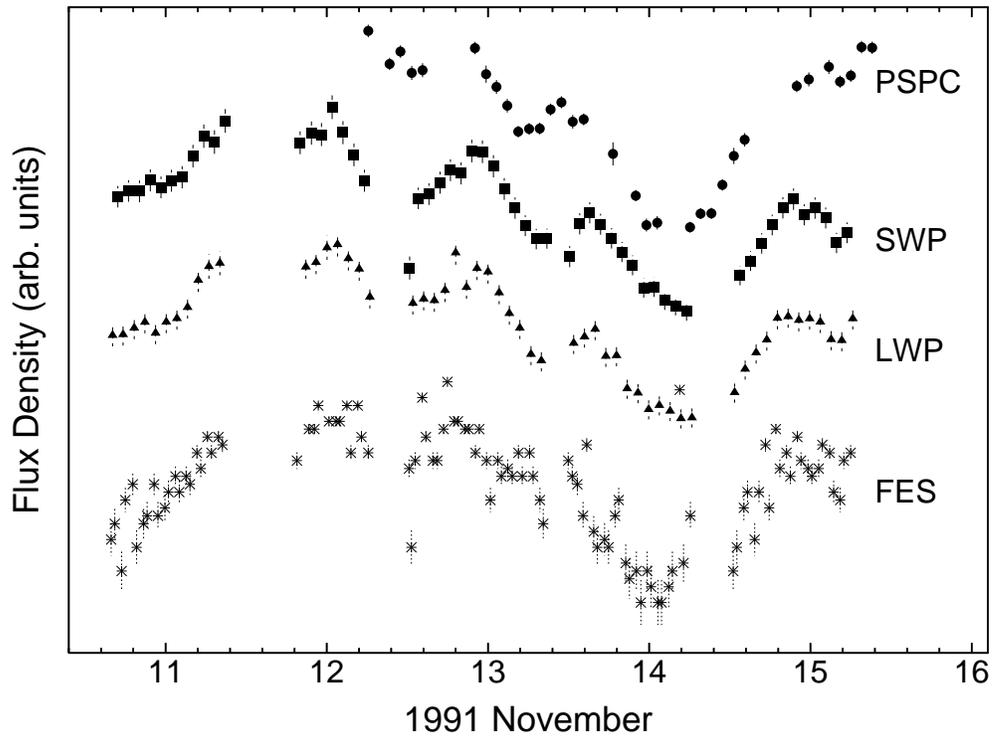}
\figcaption{Simultaneous multiwavelength flaring activity from the 1991 PKS 2155--304 campaign, adapted from \citet{ede95}, Fig. 2b. 
The \emph{IUE} FES, LWP, and SWP data are published in \citet{urr93}, the \emph{ROSAT} PSPC data are published in \citet{bri94}. 
From top to bottom, the observed wave bands are centered at 25, 1400, 2800, and 5000{\AA}.}
\end{figure}

\begin{figure}
\includegraphics[angle=270, scale=0.60]{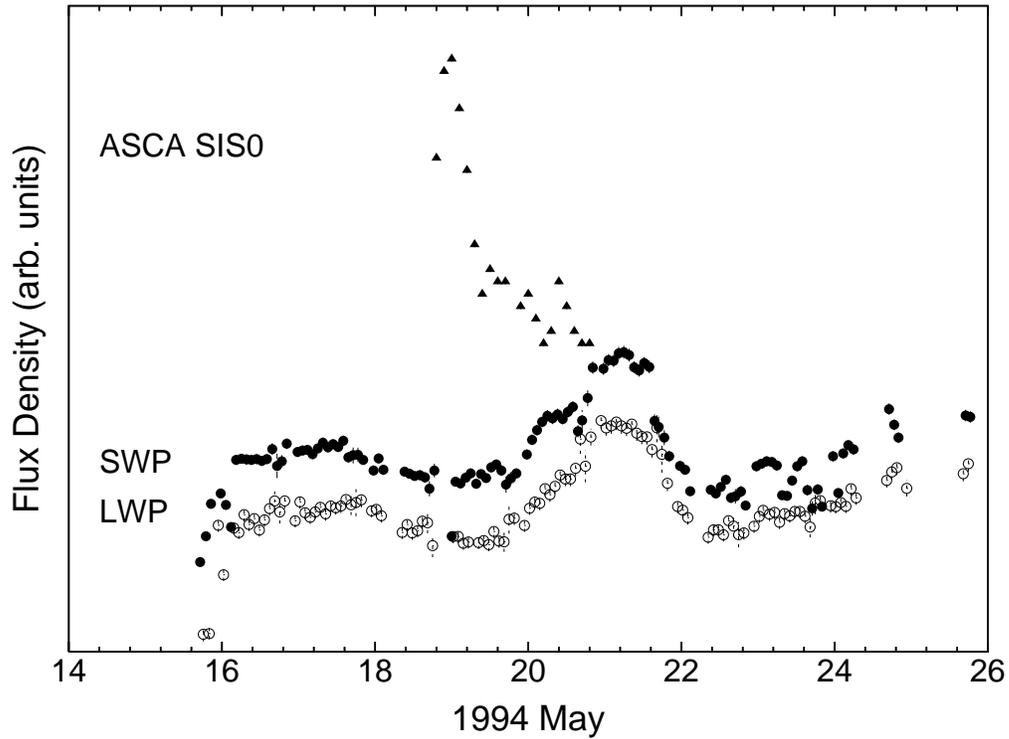}
\figcaption{Multiwavelength flaring activity from the 1994 PKS 2155--304 campaign, as discussed in \citet{urr97}. 
The \emph{IUE} SWP (1400{\AA}) and LWP (2800{\AA}) data are published in \citet{pia97}. 
The \emph{ASCA} data are binned from light-curves available on NASA's HEASARC online archives. 
The light-curves cover, from top to bottom, wave bands ranging from X-ray to UV regimes.}
\end{figure}

\begin{figure}
\includegraphics[angle=270, scale=0.50]{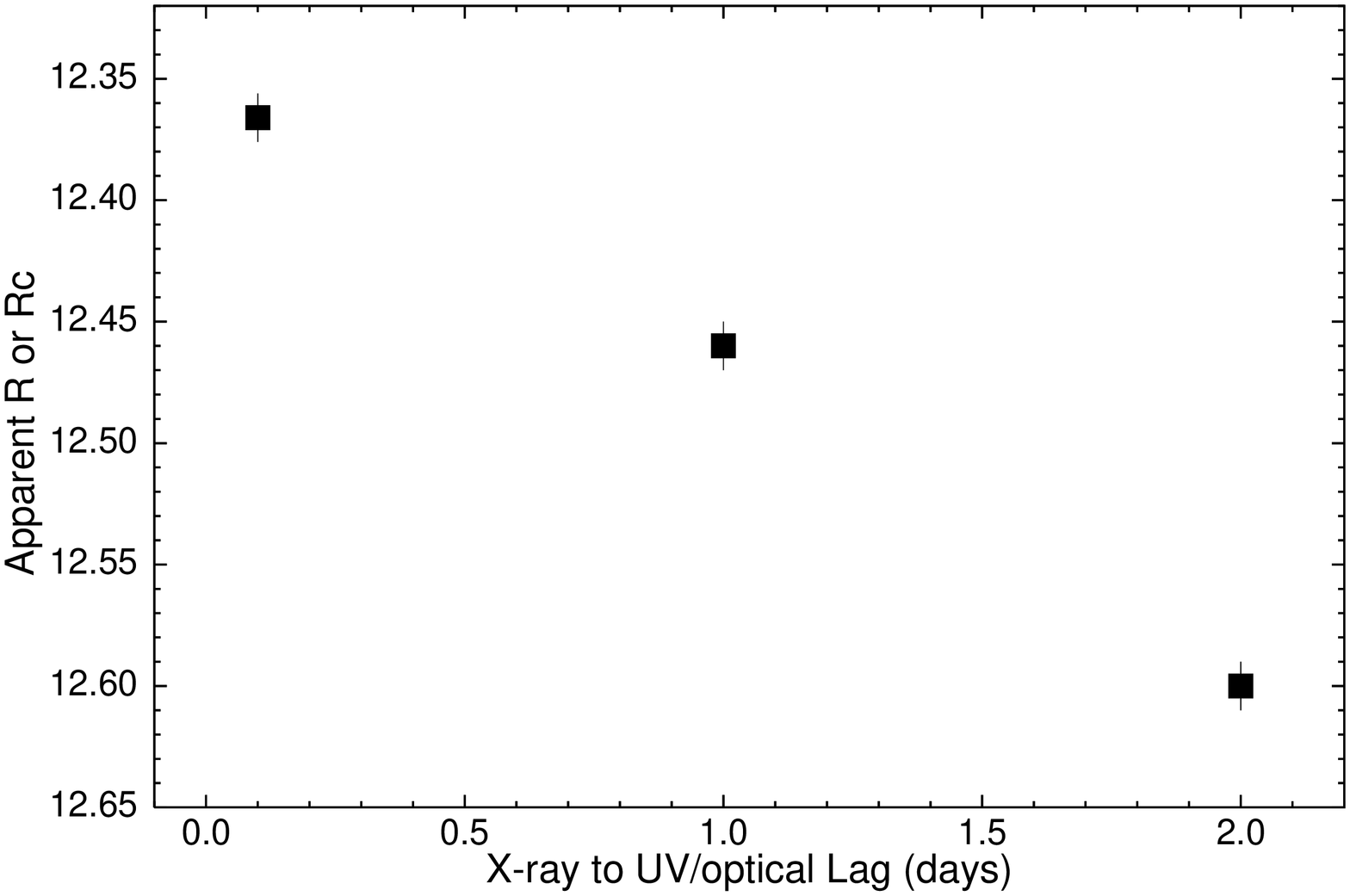}
\includegraphics[angle=270, scale=0.50]{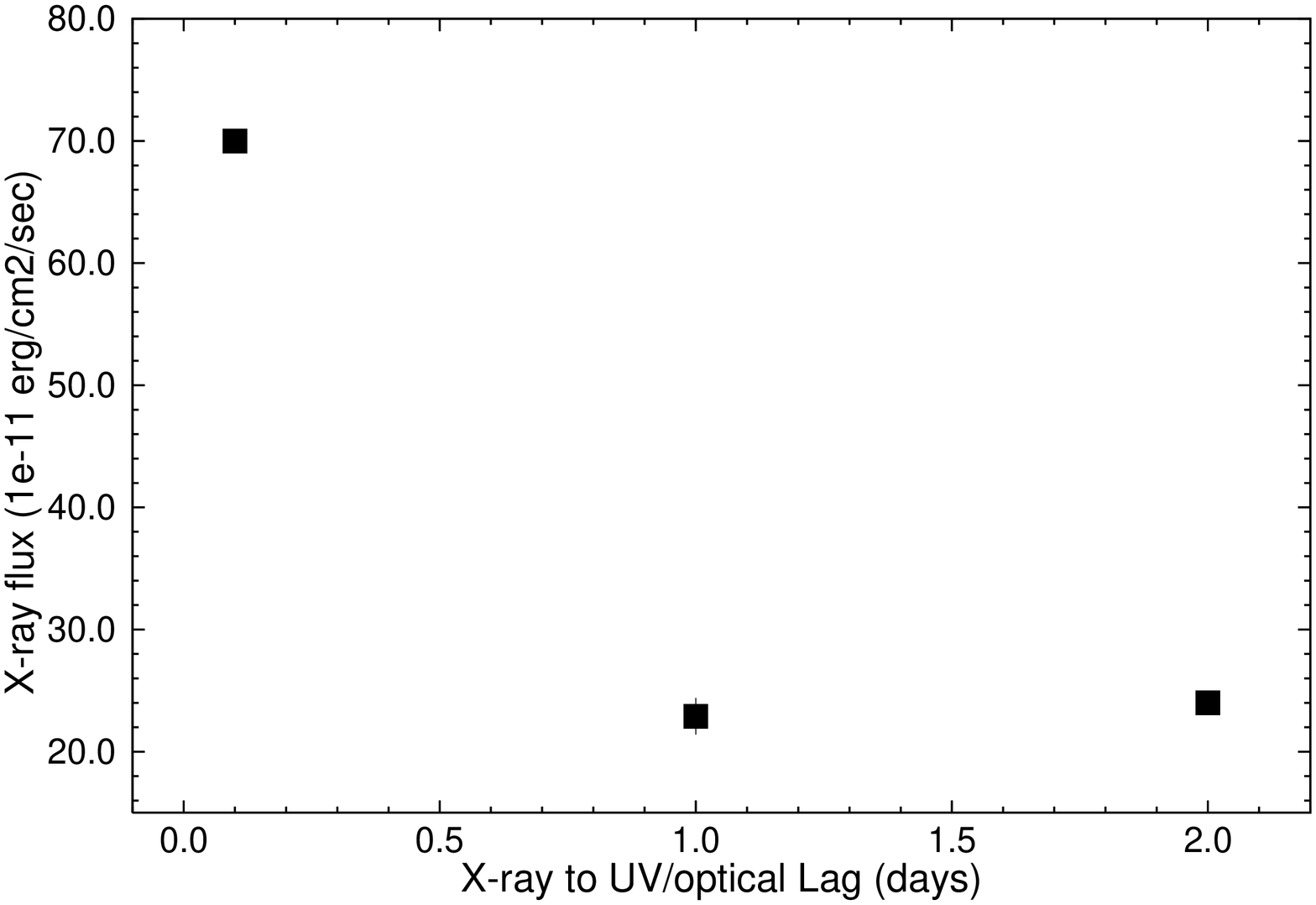}
\figcaption{The observed relationship between optical brightness (upper plot) and X-ray flux (lower plot) vs. lag time from X-ray to corresponding UV/optical activity from three simultaneous campaigns on PKS 2155--304.}
\end{figure}

\begin{figure}
\includegraphics[angle=270, scale=0.50]{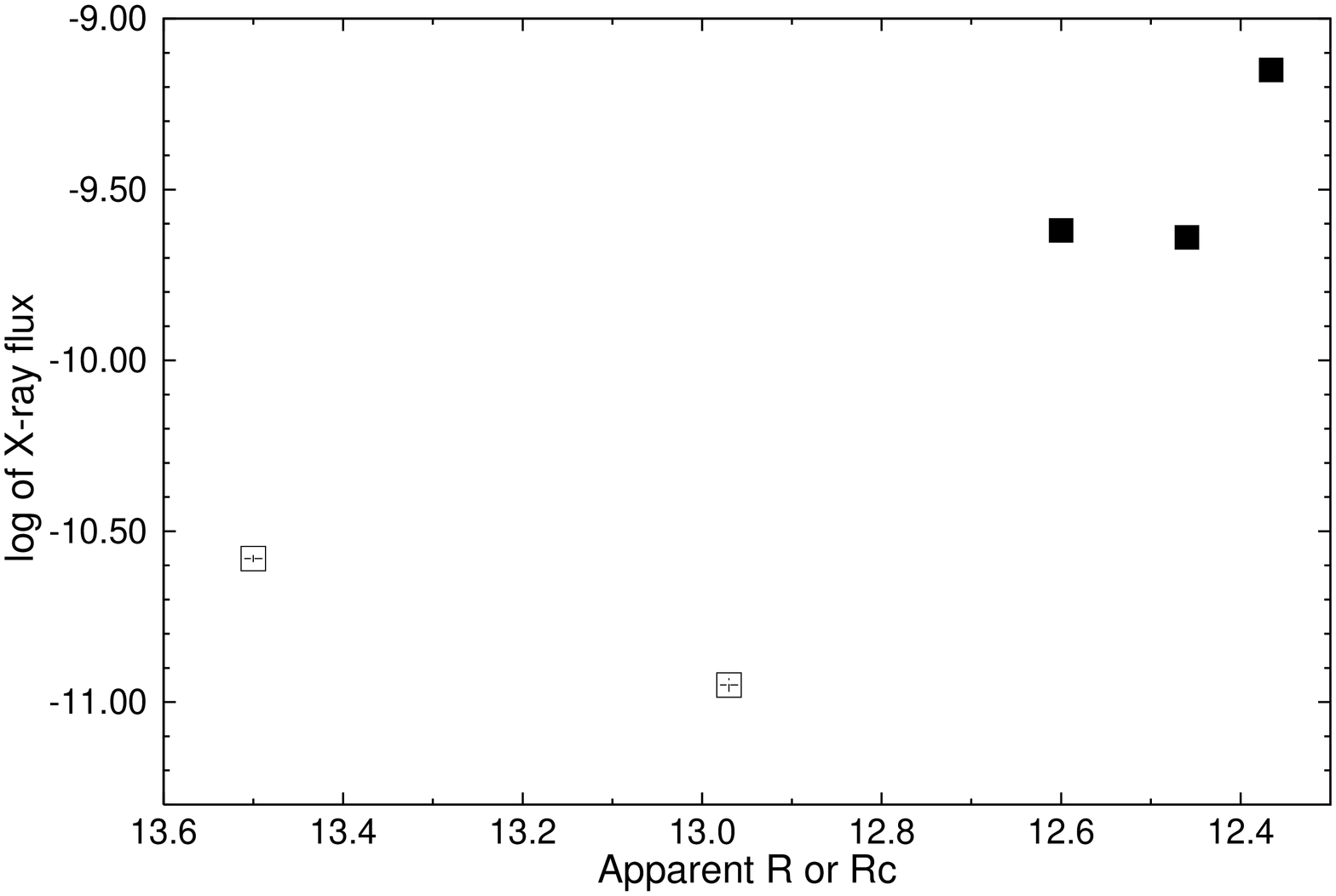}
\figcaption{The observed relationship between X-ray and optical brightness for the five PKS 2155--304 multiwavelength campaigns.
Filled squares indicate results from campaigns during which correlated X-ray and UV/optical events were observed.
Open squares indicate results from campaigns during which no correlated multiwavelength activity was observed.}
\end{figure}

\begin{figure}
\includegraphics[angle=270, scale=0.60]{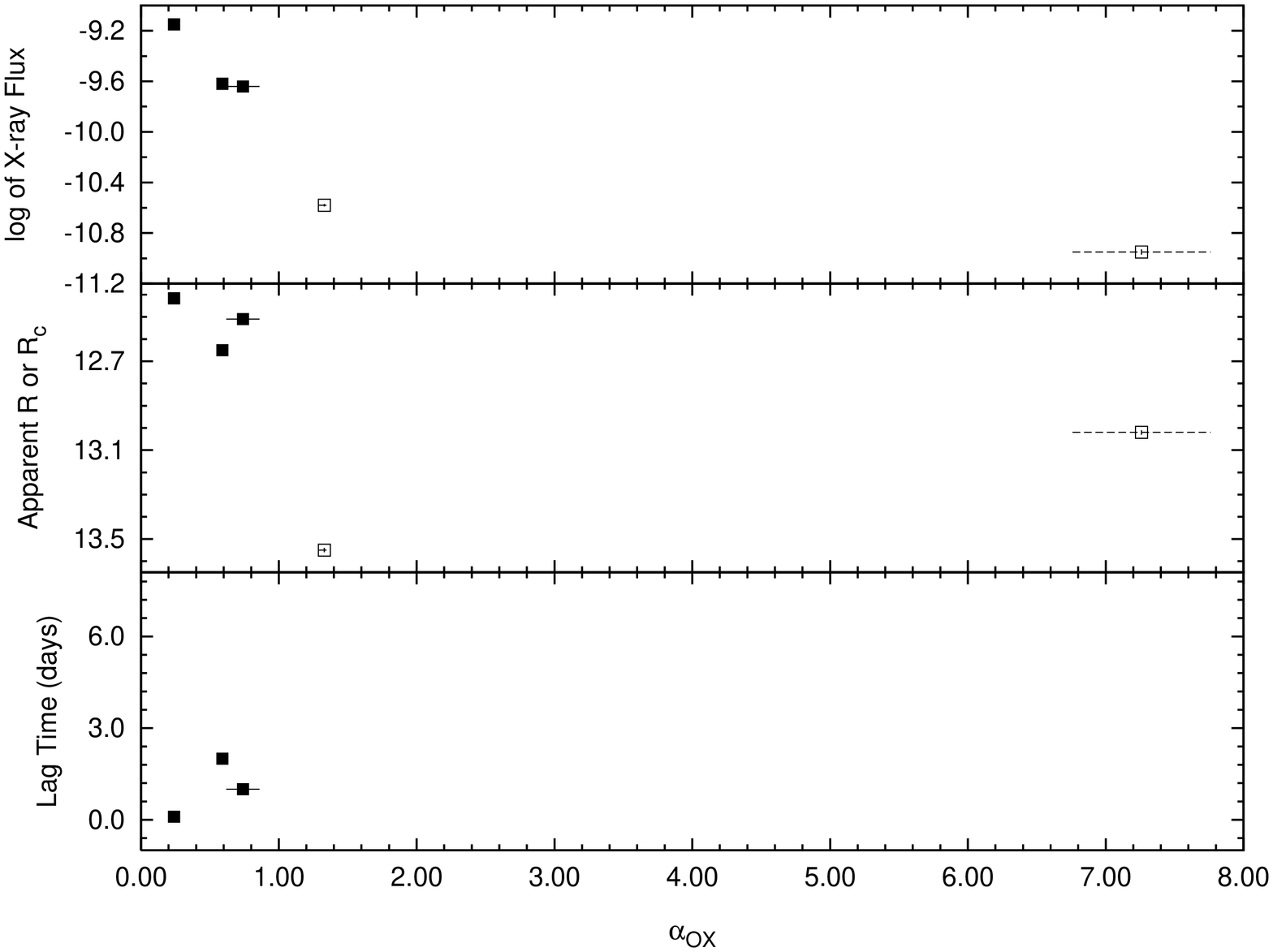}
\figcaption{The observed relationship between $\alpha_{OX}$ and lag time, X-ray brightness, and optical brightness for the five PKS 2155--304 multiwavelength campaigns.
Filled squares indicate results from campaigns during which correlated X-ray and UV/optical events were observed.
Open squares indicate results from campaigns during which no correlated multiwavelength activity was observed.}
\end{figure}

\end{document}

%% file: tab1.tex
\begin{deluxetable}{cccc}
\tablewidth{0pt}
\tablecaption{Summary of 2004 Observations.}
\tablehead{
\colhead{Observatory} & \colhead{Spectral Band} & \colhead{Dates of Obs.} & \colhead{\# of Obs.}  }
\startdata
RXTE & $2-10$ keV & 2004 Aug $5 - 25$ & 80\\
& & 2004 Sept $7 - 16$ & 20\\
CTIO (NOAO) & B-band & 2004 Aug $13 - 23$ & 36\\
& V-band & & 41\\
& R$_{c}$-band & & 1310\\
& I$_{c}$-band & & 40\\
CTIO (SMARTS) & B-band & 2004 Aug $6 - 7$ & 2\\
& V-band & 2004 Aug $6 - 7$ & 2\\
& R-band & 2004 Aug 6$ - $Oct 5 & 9\\
Lowell & R-band & 2004 Sept $22 - 23$, Oct $7 - 9$ & 8\\
UMRAO & 8.0 GHz & 2004 Sept $2 - 16$ & 5\\
& 14.5 GHz & 2004 Aug 9$ - $Sept 30 & 5\\
\enddata
\end{deluxetable}

%% file: tab2.tex
\begin{deluxetable}{ccccc}
\tablewidth{0pt}
\tablecaption{Summary of Results from PKS 2155--304 Multiwavelength Campaigns.}
\tablehead{
\colhead{Date} & \colhead{Avg. Mag.} & \colhead{Avg. X-ray Flux} & \colhead{X-ray--UV/optical Lag} & \colhead{$\alpha_{OX}$\tablenotemark{a}}  \\
& & \colhead{(10$^{-11}$ erg cm$^{-2}$ sec$^{-1}$)} & & }
\startdata
1991 Nov\tablenotemark{b} & 12.37 & $\sim$70 & 2--3 hrs & $\sim$0.24\\
1994 May\tablenotemark{c} & 12.60 & 24 $\pm$ 1 & $\sim$2 days & 0.59 $\pm$ 0.03\\
2003 Oct--Nov\tablenotemark{d} & 13.5 & 2.66 $\pm$ 0.04 & ... & 1.33 $\pm$ 0.04\\
2004 Aug\tablenotemark{e} & 12.97 & 1.11 $\pm$ 0.06 & ... & 7.26$^{+0.53}_{-0.47}$\\
2006 Jul--Aug\tablenotemark{f} & 12.46 & 22.9 $\pm$ 1.5 & $\sim$1 day & 0.74 $\pm$ 0.12\\
\enddata
\tablenotetext{a}{The ratio of the average fluxes observed in the optical and X-ray bands during each campaign.}
\tablenotetext{b}{R$_c$ mag used. Lag time from \emph{ROSAT} to \emph{IUE} wave bands. Data collected from \citet{bri94}, \citet{cou95}, and \citet{ede95}.}
\tablenotetext{c}{R$_c$ mag used. Lag time from \emph{ASCA} to \emph{IUE} wave bands. Data collected from \citet{pes97} and \citet{urr97}.}
\tablenotetext{d}{R mag used. Data collected from \citet{aha05}.}
\tablenotetext{e}{R$_c$ mag used.}
\tablenotetext{f}{R mag used. Lag time from \emph{Swift} XRT to \emph{Swift} UVOT wave bands. Data collected from \citet{fos07}.}
\end{deluxetable}